\documentstyle[12pt,graphicx]{article}
\begin{document}
\title{Vacuum Polarization by a Magnetic Flux Tube at Finite Temperature in the Cosmic String Spacetime}
\author{J. Spinelly$^{1}$ {\thanks{E-mail: jspinelly@uepb.edu.br}}  
and E. R. Bezerra de Mello$^{2}$ \thanks{E-mail: emello@fisica.ufpb.br}\\
1.Departamento de F\'{\i}sica-CCT\\
Universidade Estadual da Para\'{\i}ba\\
Juv\^encio Arruda S/N, C. Grande, PB\\
2.Departamento de F\'{\i}sica-CCEN\\
Universidade Federal da Para\'{\i}ba\\
58.059-970, J. Pessoa, PB\\
C. Postal 5.008\\
Brazil}

\maketitle
\begin{abstract}
In this paper we analyse the effect produced by the temperature in the vacuum polarization associated with charged massless scalar field in the presence of magnetic flux tube in the cosmic string spacetime. Three different configurations of magnetic fields are taken into account: $(i)$ a homogeneous field inside the tube, $(ii)$ a field proportional to $1/r$ and $(iii)$ a cylindrical shell with $\delta$-function. In these three cases, the axis of the infinitely long tube of radius $R$ coincides with the cosmic string. Because the complexity of this analysis in the region inside the tube, we consider the thermal effect in the region outside. In order to develop this analysis, we construct the thermal Green function associated with this system for the three above mentioned situations considering points in the region outside the tube. We explicitly calculate in the high-temperature limit, the thermal average of the field square and the energy-momentum tensor.
\vspace{1pc}
\end{abstract}
\maketitle

\section{Introduction}
It is well known that different types of topological defects may have been created in the early Universe after the Planck time by the vacuum phase  transition \cite{Kibble,Vilenkin}. These include domain walls, cosmic strings and monopoles. Among them cosmic string and monopole seem to be the best candidates to be detected. 

Many years ago Nielsen and Olesen proposed a theoretical model comprised by Higgs and gauge fields, that by a spontaneous broken of gauge symmetry produces linear topological defect carrying a magnetic flux named vortex \cite{N-O}. A few years later, Garfinkle investigated the influence of this topological object on the geometry of the spacetime \cite{Garfinkle}. Coupling the energy-momentum tensor associated with the system to the Einstein equations, he found static cylindrically symmetric solutions. The author also shown that asymptotically the spacetime around the vortex is a Minkowiski one minus a wedge. The core of the vortex has a non-zero thickness and the magnetic flux inside. Two years later Linet \cite{Linet} obtained, under some specific condition, exact solutions for the complete set of differential equation. He was able to show that the structure of the respective spacetime corresponds to a conical one, with the conicity parameter being expressed in terms of the energy per unity length of the vortex.\footnote{The complete analysis about the behavior of the gauge and matter fields near the cosmic string's core can only be obtained numerically. Some recent numerical analysis \cite{Yves} about the structure of supermassive cosmic strings show that two different kind of solutions for the metric tensor exist.}

In recent papers, we have investigated the vacuum polarization effects associated with massless scalar \cite{1Spinelly,Spinelly} and fermionic \cite{2Spinelly,3Spinelly} fields, in the presence of a magnetic flux tubes of finite radius in the cosmic string spacetime at zero temperature. In these analysis we considered that the magnetic fields are confined into an infinitely long tube of radius $R$ around the cosmic string. Three different configurations of magnetic field, $H(r)$, are taking into account in our analysis:   
\begin{eqnarray}
\label{h1}
i)\ \  H(r)&=&\frac{\Phi}{\alpha\pi R^2}\Theta(R-r)\ ,\ \ \mbox{homogeneous field inside,} 
\\
\label{h2} 
ii)\ \  H(r)&=&\frac{\Phi}{2\pi\alpha Rr}\Theta(R-r)\ ,\ \ \mbox{field proportional to $1/r$ inside,} 
\\
\label{h3}
iii)\ \ H(r)&=&\frac{\Phi}{2\pi\alpha R}\delta(r-R) \ , \ \ \mbox{cylindrical shell,} 
\end{eqnarray}
where $R$ is the radius extent of the tube, $\Theta$ is the Heaviside's function and $\Phi$ is the total magnetic flux.\footnote{The configurations for magnetic flux provide the same magnetic flux on the two-surface perpendicular to the $z-$axis in coordinate system defined in (\ref{1}).} The ratio of the flux to the quantum flux $\Phi_{o}$, can be expressed by $\delta=\Phi/\Phi_0=N+\gamma$, where $N$ is the integer part and $0<\gamma<1.$

In the framework of the quantum field theory at finite temperature, a fundamental quantity is the thermal Green function, $G_{\beta}(x,x')$. For the scalar field it should be periodic in the imaginary time with period $\beta$, which is proportional to the inverse of the temperature. Because we are interested to obtain the thermal Green function, it is convenient to work in the Euclidean analytic continuation of the Green function performing a Wick rotation on the time coordinate, $t\to i\tau$. So, we shall work on the Euclidean version of the idealized cosmic string spacetime, which in cylindrical coordinates, can be described  by the line element below:
\begin{equation}
ds^2=d\tau^2+dr^2+\alpha^2 r^2d\theta^2+dz^2 \ , 
\label{1}
\end{equation}
where $\alpha$ is a parameter smaller than unity which codify the presence of a conical two-surface $\left( r,\theta \right).$\footnote{For a typical Grand Unified Theory, $\alpha=1-O(10^{-6})$.}

The vacuum polarization effects associated with a charged scalar field due to a magnetic field confined in a tube of finite radius in Minkowski spacetime has been first analysed by Serebryanyi \cite{S}. A few years later this analysis, for an idealized cosmic string spacetime, has been considered by Guimar\~aes and Linet \cite{E-L}; however the magnetic flux was considered as being a line running through the string. The effect of the temperature on this vacuum polarization was also investigated by Guimar\~aes in \cite{E}. In this context, inspired by our previous work \cite{1Spinelly,Spinelly}, we decided to investigate the effect of the temperature on the vacuum polarization effects associated with charged massless scalar field in presence of magnetic flux tube in the cosmic string spacetime, considering the three different configurations of magnetic field given before. The standard procedure to develop this analysis is by calculating the respective thermal Euclidean Green function. This can be done for an ultrastatic spacetime\footnote{An ultrastatic spacetime admits a globally defined coordinate system in which the components of the metric tensor are time independent and the conditions $g_{00}=1$ and $g_{0i}=0$ hold} by knowing the Green function at zero temperature. The analysis of the thermal effects on the vacuum polarization effects associated with massless bosonic and fermionic fields in the global monopole spacetime have been considered in \cite{Mello,Mello1, Mello2} few years ago.
 
This paper is organized as follows: In the section $2$ we calculate the thermal Euclidean Green function associated with the system for the three different models of magnetic fields. Using the results obtained, we calculate in the sections $3$ and $4$, the thermal renormalized vacuum expectation value of the field square  and the energy-momentum tensor, respectively. We leave for the section $4$ our conclusions.

\section{The Euclidean thermal Green function}
\label{sec2}

The Green function associated with the charged massless scalar field at zero temperature in the presence of a electromagnetic field, must obey the following non-homogeneous second-order differential equation
\begin{equation}
\frac{1}{\sqrt{-g}}D_{\mu}\left[\sqrt{-g}g^{\mu\nu} D_{\nu}\right]G(x,x^{'})=-\delta^{(4)}(x-x') \ ,
\label{j5}
\end{equation} 
where $D_{\mu}=\partial_\mu-ieA_\mu$, $A_\mu$ being the four-vector potential. 

In order to reproduce the configurations of magnetic fields along the $z-$direction given by (\ref{h1})-(\ref{h3}), we write the the vector potential by $A_{\mu}=(0,0,A(r),0)$, with
\begin{equation}
A(r)=\frac{\Phi}{2\pi}a(r) \ .
\label{j8} 
\end{equation}

For the two first models considered, we can represent the radial function $a(r)$ by:
\begin{equation}
a(r)=f(r)\Theta (R-r)+\Theta (r-R) \ ,
\label{j9}
\end{equation}
with
\begin{eqnarray}
f(r)=\left\{\begin{array}{cc}
r^2/R^2,&\mbox{for the model ({\it{i}}) and}\\
r/R,&\mbox{for the model ({\it{ii}}).}
\end{array}
\right.
\label{j10}
\end{eqnarray}
For the third model,
\begin{equation}
\label{jj10}
a(r)=\Theta(R-r) \ .
\end{equation}

As we have mentioned, in this work we shall continue in the same line of investigation started in \cite{1Spinelly,Spinelly}, calculating at this time, the thermal contribution on the vacuum polarization effects in the region outside the magnetic tube for the three magnetic fields under consideration. 

The Euclidean Green functions for points outside the magnetic flux tube at zero-temperature are given below:
\begin{itemize}
\item For the models $(i)$ and $(ii)$,
\begin{eqnarray}
G^{j}_{T=0}(x,x')&=&\frac{e^{i N\Delta\theta}}{8\pi^2\alpha rr'\sinh u_0}\frac{e^{i\Delta \theta}\sinh(\gamma u_0/\alpha)+\sinh[(1-\gamma)u_0/
\alpha]}{\cosh(u_0/\alpha)-\cos\Delta\theta}
\nonumber\\
&+&\frac{1}{4 \pi^{2} \alpha}\int_0^{\infty} d\omega \omega J_{0}\left(\omega\sqrt{(\Delta 
\tau)^{2}+(\Delta z)^{2}}\right)\times
\nonumber\\
&&\sum_{n=-\infty}^{\infty}e^{i n\Delta\theta}D^j_n(\omega R)K_{|\nu|}(\omega r)K_{|\nu|}(\omega r') , \qquad{j=1,2} \ ,
\label{j19}
\end{eqnarray}
where 
\begin{eqnarray}
\nu =\frac{n-\delta}\alpha=\frac{n-N-\gamma} \ ,
\end{eqnarray}
\begin{equation}
\cosh u_{o}=\frac{r^2+{r^{'}}^{2}+(\Delta \tau)^{2}+(\Delta z)^{2}}{2rr^{'}} 
\label{j20}
\end{equation}
and
\begin{equation}
D_n^j(\omega R)=\frac{H_j'(R)I_{|\nu|}(\omega R)-H_j(R)I'_{|\nu|}(\omega R)}
{H_{j}(R)K^{'}_{|\nu|}(\omega R)-H_j'(R)K_{|\nu|}(\omega R)} \ .
\label{j17}
\end{equation} 
In the above equations the functions $H_j(r)$ are given by:
\begin{equation}
H_{1}(r)=\frac{1}{r}M_{\sigma_{1}, \lambda_{1}}\left( \frac{\delta}{\alpha R^2} 
r^2\right),
\label{j15}
\end{equation}
with $\sigma_{1}=(\frac{n}{\alpha}-\frac{\omega^2 R^2\alpha}{2\delta})/2$ and
$\lambda_{1}=n/2\alpha$, and 
\begin{equation}
H_{2}(r)=\frac{1}{\sqrt{r}}M_{\sigma_{2}, \lambda_{2}}\left( \zeta r\right),
\label{j16}
\end{equation}
with $\sigma_{2}=\frac{n\delta}{\alpha}(\delta^2+\omega^2 R^2 \alpha^2)^{-1/2}$, $\lambda_{2}=n/\alpha$ and $\zeta=\frac{2}{R\alpha}(\delta^2+\omega^2 R^2 \alpha^2)^{1/2}$. Moreover, $M_{\sigma, \lambda}$ is the Whittaker function, while $I_{|\nu|}$ e $K_{|\nu|}$ are the modified Bessel functions \cite{Grad}.
\item For the model $(iii)$, we have
\begin{eqnarray}
G^{(3)}_{T=0}(x,x')&=&\frac1{8\pi^2 \alpha rr'\sinh u_0}\left[\frac{\sinh(u_0/\alpha)}{\cosh(u_{o}/\alpha)-\cos\Delta\theta}\right]
\nonumber\\
&+&\frac1{4 \pi^{2}\alpha}\int_0^{\infty} d\omega \omega J_{0}\left(\omega\sqrt{(\Delta \tau)^{2}+(\Delta z)^{2}}\right)\times
\nonumber\\
&&\sum_{n=-\infty}^\infty e^{in\Delta\theta}D_n(\omega R)K_{|n/\alpha|}(\omega r)K_{|n/\alpha|}(\omega r') \ .
\label{24}
\end{eqnarray}
where
\begin{equation}
D_{n}(\omega R)=\frac{I_{|\nu|}^{'}(\omega R)I_{|n|/\omega}(\omega R)-I_{|\nu|}(\omega R)I^{'}_{|n|/\alpha}(\omega R)}{I_{|\nu|}(\omega R)K^{'}_{|n|/\alpha}(\omega R)-I_{|\nu|}^{'}(\omega R)K_{|n|/\alpha}(\omega R)} \ .
\end{equation}
\end{itemize}

We can observe that the first term on the right hand side of (\ref{j19}) is, up to a gauge transformation, equivalent to the result presented by Guimar\~aes and Linet \cite{E-L} for a charged massless scalar field in the presence of a magnetic flux running through the cosmic string; and as to (\ref{24}), its first term corresponds the results found by Smith \cite{smith} and by Linet in \cite{linet2} for a massless scalar field without charge. However, the seconds terms of booths expressions represent corrections on the respective Green functions due to a non-vanishing radius $R$ attributed to the magnetic flux; off course these corrections vanish when we take $R\to 0$.

Following the prescription given in the papers by Braden \cite{Braden} and Page \cite{Page}, the thermal Green function, $G_{T}(x,x')$, can be expressed in terms of the sum
\begin{equation}
G_{T}(x,x')=\sum_{l=-\infty}^{\infty}G_{T=0}(x,x'-l\lambda\beta) \ , 
\label{temp}
\end{equation} 
where $\lambda$ is the "Euclidean" unitary time-like vector and $\beta=1/k_BT$, being $k_B$ the Boltzmann constant and $T$ the absolute temperature. 

In agreement with the equations (\ref{j19}), (\ref{24}) and (\ref{temp}), the thermal Green functions associated with the massless scalar field in the cosmic string spacetime and in the presence of magnetics field, are given by: 
\begin{itemize}
\item For the models $(i)$ and $(ii)$,
\begin{eqnarray}
{G}^{j}_{T}(x,x')&=&G_T^\gamma(x,x')+\frac{1}{4 \pi^{2} \alpha}\sum_{l = -\infty}^\infty\int_0^{\infty} d\omega \omega J_{0}\left(\omega\sqrt{(\Delta 
\tau+l\beta)^{2}+(\Delta z)^{2}}\right)\times
\nonumber\\
&&\sum_{n=-\infty}^{\infty}e^{i n\Delta\theta}D^j_n(\omega R)K_{|\nu|}(\omega r)K_{|\nu|}(\omega r') \ , \qquad{j=1,2} \ .
\label{j0}
\end{eqnarray}
\item For the model $(iii)$,
\begin{eqnarray}
{G}^{(3)}_{T}(x,x')&=&G_T(x,x')+\frac{1}{4 \pi^{2} \alpha}\sum_{l = -\infty}^{\infty}\int_0^{\infty} d\omega \omega J_{0}\left(\omega\sqrt{(\Delta \tau+l\beta)^{2}+(\Delta z)^{2}}\right)\times
\nonumber\\
&&\sum_{n=-\infty}^{\infty}e^{i n\Delta\theta}D_n(\omega R)K_{|n|/\alpha}(\omega r)K_{|n|/\alpha}(\omega r') \ .
\label{jj1}
\end{eqnarray}
\end{itemize}

The thermal Green functions $G_T^\gamma(x,x')$ and $G_T(x,x')$ which appear in (\ref{j0}) and (\ref{jj1}) have being obtained a few years ago by Guimar\~aes \cite{E} and Linet \cite{linet1}, respectively. So, we shall not repeat them. In fact what we are really interested in this paper is to analyze the seconds terms in these Green functions, which correspond to the thermal contributions due to the finite thickness admitted for the magnetic flux tube.

\section{The Computation of $\langle\hat{\phi}^{\ast}(x)\hat{\phi}(x)\rangle_{ Ren.}$ at Non-zero Temperature}
\label{sec3}

The main objective of this section is to investigate the effects produced by the temperature in the renormalized vacuum expectation value of the square of the charged massless scalar field, $\langle\hat{\phi}^{\ast}(x)\hat{\phi}(x)\rangle$, in the presence of a magnetic flux tube of finite radius. Formally this quantity is given by taking the coincidence limit of the Green function:
\begin{eqnarray}
\langle\hat{\phi}^{\ast}(x)\hat{\phi}(x)\rangle_{T}=\lim_{x'\rightarrow x}G_{T}(x,x')	 \ .
\end{eqnarray}
However, this procedure provides a divergent result and the divergence comes exclusively from the first terms of the right hand side of ({\ref{j0}) and (\ref{jj1})\footnote{A special feature of these Green functions is that the correction due to the magnetic tube's radius is finite in the coincidence limit.}. In order to obtain a finite and well defined result, we must apply some renormalization procedure. Here we shall adopt the point-splitting renormalization one. It has been observed that the singular behavior of the Green function has the same structure as given by the Hadamard one, which on the other hand can be written in terms of the square of the geodesic distance between two points. So, here we shall adopt the following prescription: we subtract from the Green function the Hadamard one before applying the coincidence limit as shown below:
\begin{eqnarray}
\langle\hat{\phi}^{\ast}(x)\hat{\phi}(x)\rangle_{T, Ren.}=\lim_{x'\rightarrow x}\left[G_{T}(x,x')-G_{H}(x,x')\right] \ .
\end{eqnarray}
We can write the result as:
\begin{eqnarray}
\langle\hat{\phi}^{\ast}(x)\hat{\phi}(x)\rangle_{T, Ren.}=\langle\hat{\phi}^{\ast}(x)\hat{\phi}(x)\rangle_{T, Reg.}+ \langle\hat{\phi}^{\ast}(x)\hat{\phi}(x)\rangle_{T=0}^{C} +\langle\hat{\phi}^{\ast}(x)\hat{\phi}(x)\rangle_{T}^{C} \ . 
\label{new}
\end{eqnarray}

The first term on the right hand side of the above expression, represents, for the models $(i)$ and $(ii)$, the thermal contribution coming from the interaction between charged massless scalar field with a magnetic flux considered as a line running along the cosmic string. From the Guimar\~aes's paper \cite{E}, this term is: 
\begin{eqnarray}
\langle\hat{\phi}^{\ast}(x)\hat{\phi}(x)\rangle_{T, Reg.}=\frac{1}{12 \beta^2}+\frac{1}{16\pi^{2}\alpha\beta r}\int_{0}^{\infty}\frac{\coth \left[ \frac{2\pi}{\beta}r \cosh u/2\right]}{\cosh u/2}F_{\alpha}^{(\gamma)}\left(u,0 \right) \ du \ ,
\label{old}
\end{eqnarray}
where
\begin{eqnarray}
F_{\alpha}^{(\gamma)}\left(u,0 \right)=-2\frac{\sin\left[ \pi\gamma/\alpha\right]\cos\left[ u\left(1-\gamma \right)/\alpha\right]+\sin\left[ u\left(1-\gamma \right)/\alpha\right]\cos\left[ \pi\gamma/\alpha\right]}{\cosh u/\alpha-\cos\pi/\alpha} \ . 
\end{eqnarray}
For the model $(iii)$ an analogous expression can be obtained from the previous one, by taking $\gamma=0$\footnote{See paper \cite{linet1}.}. (An interesting aspect of these results is that the vacuum polarizations depend only on the fractional part of the ration of the magnetic flux by the quantum one, $\gamma$.)

In the high-temperature limit ($\beta\to 0$), Guimar\~aes showed that,
\begin{equation}
\langle\hat{\phi}^{\ast}(x)\hat{\phi}(x)\rangle_{T, Reg.}\approx \frac{1}{12 \beta^{2}}+\frac{M^{(\gamma)}}{\beta r} \ ,
\end{equation}
where the constant $M^{(\gamma)}$ is given by
\begin{eqnarray}
M^{(\gamma)}=\frac{1}{16 \pi^{2} \alpha}\int_{0}^{\infty}\frac{F_{\alpha}^{(\gamma)}\left( u,0\right)}{\cosh(u/2)} \ du \ ,
\end{eqnarray}
and in the zero-temperature limit ($\beta\to\infty$),
\begin{equation}
\langle\hat{\phi}^{\ast}(x)\hat{\phi}(x)\rangle_{T, Reg.}\approx \frac{\omega(\gamma)}{2r^2} \ ,
\end{equation}
with $\omega{(\gamma)}$ being given by
\begin{eqnarray}
\omega(\gamma)=-\frac1{8\pi^2}\left\{\frac13-\frac1{2\alpha^2}\left[4\left(\gamma-\frac12\right)^2-\frac13\right]\right\} \ .
\end{eqnarray}

The two last terms in (\ref{new}) are corrections on the vacuum polarization due to finite thickness of the radius of the magnetic tube and non-zero temperature. The term $\langle\hat{\phi}^{2}(x)\rangle_{T=0}^{C}$, corresponds to correction due to finite thickness of the radius of tube only, it was given in \cite{1Spinelly,Spinelly}. For the two firsts models the corrections are similar. They are given by the component $l=0$ in (\ref{j0}), and read
\begin{eqnarray}
\frac1{(2\pi)^{2}\alpha}\int_0^{\infty} d\omega \omega \sum_{n=-\infty}^\infty D^i_n(\omega R)K^{2}_{|\nu|}(\omega r) \ . 
\label{correc}
\end{eqnarray}
Because we are mostly interested to study the vacuum polarization for points very far from the cosmic string, we shall consider $R/r\ll 1 $. Comparing  the behavior of $D^i_n(\omega R)$ with the behavior of $K^{2}_{|\nu|}(\omega r)$, we have shown that we may approximate the integrand of (\ref{correc}) assuming to the coefficient $D_n^j$ its first order expansion in $\omega R$ , which reads:
\begin{equation}
D^j_n(\omega R)=-\frac{2}{\Gamma(|\nu|+1)\Gamma(|\nu|)}\left(\frac{w^n_j-|\nu|z^{n}_{j}}{w^{n}_{j}+|\nu|z^{n}_{j}}\right)\left( \frac{\omega R}{2} \right)^{2|\nu|},
\label{27}
\end{equation} 
where
\begin{equation}
w^{n}_{1}=\left( \frac{\delta-n}{\alpha}-1\right)M_{\lambda_1,\lambda_2}(\delta/\alpha)+\left( 1+\frac{2n}{\alpha}\right)M_{\gamma_1,\lambda_1} (\delta/\alpha),
\label{28}
\end{equation}
\begin{equation}
z^{n}_{1}=M_{\lambda_{1},\lambda_{2}}(\delta/\alpha),
\end{equation}
\label{29}
\begin{equation}
w^{n}_{2}=\left( \frac{\delta-n}{\alpha}-\frac{1}{2}\right)M_{\lambda_2, \lambda_1}(2\delta/\alpha)+\left(\frac12+\frac{2n}{\alpha}\right)M_{\gamma_2,
\lambda_2}(2\delta/\alpha) \ ,
\label{30}
\end{equation}
and
\begin{equation}
z^n_2=M_{\lambda_2,\lambda_1}(2\delta/\alpha) \ ,
\label{31}
\end{equation}
being $\gamma_1=(n+2\alpha)/2\alpha$, $\gamma_2=(n+\alpha)/\alpha$ and $\nu=\frac{n-N-\gamma}\alpha.$ The most important contributions in (\ref{correc}) comes from the component $n=N$ in the summation. In this way, considering only the dominant term, the corrections are given by:
\begin{eqnarray}
\langle\hat{\phi}^{\ast}(x)\hat{\phi}(x)\rangle_{T=0}^{C}=-\frac{1}{(2\pi)^2r^2}\frac{\gamma}{\alpha(2\gamma+\alpha)}\left[\frac{w^N_j-(\gamma/\alpha)z^N_j}{w^N_j+(\gamma/\alpha)z^N_j}\right]
\left(\frac{R}{r} \right)^{2\gamma/\alpha} \ .
\label{j33}
\end{eqnarray}
As we can see the corrections present extra dependence on the radial coordinate, consequently they are appreciable only in the region close to the tube. 

As to the third model there happens a very interesting phenomenon. For $n \not= 0$, the coefficient $D_n(\omega R)$ in (\ref{24}) vanishes at least as fast as $(\omega R)^{2|n|/\alpha}$ when $\omega R \rightarrow 0$. On the other hand, $D_{0}(\omega R)$ vanishes only with 
the inverse of the logarithm, so quite slowly. In this way the most relevant contribution to the summation comes from $n=0$. For this case 
\begin{equation}
D_0(\omega R)=\frac1{\ln\left(\frac{\omega R}2\right)+{\mathcal{C}}-\alpha/\delta} \ ,
\label{37}
\end{equation}
where ${\mathcal{C}}$ is the Euler constant. In \cite{Spinelly} we have explicitly shown that the correction on the vacuum polarization effect for the third model, consequence of a non-vanishing radius of the magnetic flux, is mainly given by:
\begin{equation}
\label{p2}
\langle\hat{\phi}^{\ast}(x)\hat{\phi}(x)\rangle_{T=0}^{C}=-\frac1{8\pi^2\alpha r^2\ln\left(\frac{2r}R e^{-{\mathcal{C}}+\alpha/\delta}\right)} \ .
\end{equation}
We also have shown that for $\alpha=0.99$ and $\delta=0.2$ this term is of the same order of the standard vacuum expectation value of the field square in the absence of magnetic flux, up to the distance $r$ which exceeds the radius of the observable Universe.

As we have already mentioned, the last term in (\ref{new}), $\langle\hat{\phi}^{\ast}(x)\hat{\phi}(x)\rangle_{T}^{C}$, is consequence of combined effects of the nonvanishing thickness of the magnetic tube and temperature. This is a new contribution and we shall focus on it. For the first two models, this term can be expressed by
\begin{equation}
\langle\hat{\phi}^{\ast}(x)\hat{\phi}(x)\rangle_{T}^{C}=\frac{1}{2\pi^{2}\alpha r^{2}}\sum_{l=1}^{\infty}\int_0^{\infty} dv v J_{0}\left(v\zeta l\right)\sum_{n=-\infty}^{\infty}D^j_n(vR/r)K^{2}_{|\nu|}(v)\  , \quad{j=1,2}  \ ,
\label{23}
\end{equation}
with $\zeta=\beta/r$. In the above expression we have introduced a dimensionless variable $v=\omega r$.

As in the zero temperature analysis, because we are considering $R/r\ll 1$, the most important contribution for the above summation comes from the $n=N$ component. So, on basis what we have already discussed for points very far from the string, equation (\ref{23}) can be approximate to
\begin{eqnarray}
\label{PP}
\langle\hat{\phi}^{\ast}(x)\hat{\phi}(x)\rangle_{T}^{C}&=&-\frac{1}{(2\pi)^{2}\alpha r^{2}}\frac{2} {\Gamma(\gamma/\alpha+1) \Gamma(\gamma/\alpha)}\left(\frac{w^N_j-(\gamma/\alpha)z^{N}_{j}}{w^{N}_{j}+(\gamma/\alpha)z^{N}_{j}}\right)\left( \frac{R}{2r}\right)^{2\gamma/\alpha} \times \nonumber \\
&& \sum_{l=1}^{\infty}\int_0^{\infty} dv v^{1+2\gamma/\alpha} J_{0}\left(v\zeta l\right)K^{2}_{|\gamma/\alpha|}(v)  \  , \quad{j=1,2} \ . 
\end{eqnarray}
 
From the above expression we can observe that the thermal content of this correction is in the summation $S$ given below: 
\begin{eqnarray}
S(\zeta)=\sum_{l=1}^{\infty}\int_0^{\infty} dv v^{1+2\gamma/\alpha} J_{0}\left(v\zeta l\right)K^{2}_{|\gamma/\alpha|}(v) \ .
\label{Sum}
\end{eqnarray} 
Unfortunately it is not possible to obtain a closed expression to this term, and provide a complete information about the thermal behavior of (\ref{PP}); on the other hand, it is possible to give its main information. In order to do that we shall divide the integral above in two parts: from $\left[0,2\pi/\zeta \right]$ and from $\left[2\pi/\zeta,\infty \right)$. Due the strong exponential decay of the modified Bessel function for large argument, the contribution in the interval $\left[2\pi/\zeta,\infty \right)$ can be neglected in the high-temperature regime, i.e., for $\zeta\ll 1$; moreover using the series properties for the Bessel function \cite{Grad}, 
\begin{equation}
\sum_{l=1}^{\infty}J_{0}(v\zeta l)=-\frac{1}{2}+\frac{1}{\zeta v} \ , \quad 0<v<2\pi/\zeta \ ,
\end{equation} 
(\ref{Sum}) can be approximated to
\begin{eqnarray}
S(\zeta)=-\frac{1}{2}\int_0^{2\pi/\zeta} dv v^{1+2\gamma/\alpha}K^{2}_{|\gamma/\alpha|}(v)+\frac{1}{\zeta}\int_0^{2\pi/\zeta} dv v^{2\gamma/\alpha} K^{2}_{|\gamma/\alpha|}(v)  \ .
\end{eqnarray} 
In the high-temperature limit the most relevant contribution is given by the second term. Adopting the same approximation criterion to discard the integral in the interval $\left[2\pi/\zeta,\infty \right)$, we may evaluate the integral by taking its upper limit going to infinity. In this way the most relevant contribution to the thermal vacuum polarization effect is \cite{Grad}\footnote{By numerical analysis, in \cite{Spi} we have reached to $S(\zeta)$, the same behavior found here, i.e., $S(\zeta)\approx 1/\zeta$.}:
\begin{equation}
S(\zeta)=\frac{1}{\zeta}\frac{\sqrt{\pi}}{2^{2-2\gamma/\alpha}}\frac{\Gamma^{2}\left( 1/2+\gamma/\alpha\right)\Gamma\left(1/2+2\gamma/\alpha \right)}{\Gamma\left(1+2\gamma/\alpha \right)} \ .
\end{equation} 
Consequently, in the high-temperature regime the correction term, $\langle\hat{\phi}^{\ast}(x)\hat{\phi}(x)\rangle_{T}^{C}$, is dominated by:
\begin{equation}
\langle\hat{\phi}^{\ast}(x)\hat{\phi}(x)\rangle_{T}^{C}=-\frac{A_j^{(\gamma, \alpha, R)}}{ r \beta}\left( \frac{R}{r}\right)^{2\gamma/\alpha}  \  , \quad{j=1,2}  \ ,
\end{equation}
with
\begin{eqnarray}
A_j^{(\gamma, \alpha, R)}=\frac{1}{4\pi^{3/2}\alpha}\frac{\Gamma^{2}\left( 1/2+\gamma/\alpha\right)\Gamma\left(1/2+2\gamma/\alpha \right)}{\Gamma\left(1+2\gamma/\alpha \right)\Gamma(1+\gamma/\alpha)\Gamma(\gamma/\alpha)}\left[\frac{w^N_j
-(\gamma/\alpha)z^{N}_{j}}{w^{N}_{j}+(\gamma/\alpha)z^{N}_{j}}\right] \ . \nonumber
\end{eqnarray}
 
Hence, the expression for renormalized vacuum expectation value of the square of the charged massless scalar field, in the high-temperature limit, is given by
\begin{equation}
\langle\hat{\phi}^{\ast}(x)\hat{\phi}(x)\rangle_{T, Ren.}\approx \frac{1}{12 \beta^{2}}+\frac{M^{(\gamma)}}{\beta r} -\frac{A_j^{(\gamma, \alpha, R)}}{r \beta}\left( \frac{R}{r}\right)^{2\gamma/\alpha} \ .
\end{equation}
From this expression we can observe that the two sub-dominant contributions are of the same order of magnitude for points near the tube.

For the third model, $\langle\hat{\phi}^{\ast}(x)\hat{\phi}(x)\rangle_{T}^{C}$, is given by the expression below: 
\begin{equation}
\langle\hat{\phi}^{\ast}(x)\hat{\phi}(x)\rangle_{T}^{C}=\frac{1}{2\pi^{2}\alpha r^{2}}\sum_{l=1}^{\infty}\int_0^{\infty} dv v J_{0}\left(v\zeta l\right)\sum_{n=-\infty}^{\infty}D_n(vR/r)K^{2}_{|n|/\alpha}(v) \  .
\label{j1}
\end{equation}

Again, as in the zero temperature analysis of the vacuum polarization effect, the most relevant contribution comes from the $n=0$ component of the above summation for $R/r\ll 1$. The respective coefficient is:
\begin{equation}
D_0(vR/r)=\frac{1}{\ln\left( v/qr\right)} \ ,
\label{ap}
\end{equation}
where
\begin{equation}
q=\frac{2}{R}e^{-\mathcal{C}+\alpha/\delta} \ .
\end{equation}

In this way (\ref{j1}) can be written by:
\begin{equation}
\langle\hat{\phi}^{\ast}(x)\hat{\phi}(x)\rangle_{T}^{C}=\frac{1}{2\pi^{2}\alpha r^{2}}\bar{S}(\zeta)\  ,
\end{equation}
where
\begin{equation}
\bar{S}(\zeta)=\sum_{l=1}^{\infty}\int_0^{\infty} dv v  J_{0}\left(v\zeta l\right)D_0(vR/r) K^{2}_{0}(v) \ .
\label{j2}
\end{equation}

Applying the previous summation property to the Bessel functions, in the high-temperature regime, (\ref{j2}) can be given by
\begin{equation}
\bar{S}(\zeta)=-\frac{1}{2}\int_0^{2\pi/\zeta} dv v D_0(vR/r) K^{2}_{0}(v)+\frac{1}{\zeta}\int_0^{2\pi/\zeta} dv   D_0(vR/r)K^{2}_{0}(v) \ .
\label{j3}
\end{equation}

For $\zeta<<1$ the most relevant contribution to $\bar{S}$ is given by second term; moreover in this limit we can obtain an approximated expressions for it by taking the upper limit of the integral going to infinite. In this way (\ref{j3}) can be evaluated by
\begin{equation}
\bar{S}(\zeta)\cong\frac{1}{\zeta}\int_0^{\infty} dv   \frac{K^{2}_{0}(v)}{\ln(v)-\ln(qr)} \ .
\label{jj3}
\end{equation}

Unfortunately this expression presents a pole for $v=qr$. However, this pole is consequence of the approximation adopted and it is located in the region where where the approximation is no more valid\footnote{For $\gamma=0.2$, $N=0$ and $\alpha=0.99$, $qr=\frac{2r}Re^{{\cal{C}}-\alpha/\zeta}$ is of order $10^5$ for $r/R=10^3$, so $vR/r=10^2$ consequently bigger than unity.}. In fact the full expression to $D_0(vR/r)$ has no singularity. $D_0$ is a slowly increasing function of $v$. As we have mentioned, the full expression to the coefficient $D_0$ vanishes with the inverse of the logarithm for $v\approx 0$, and grows less slower than $e^{2vR/r}$ for large $v$. The square of the modified Bessel function, $K_0^2$, provides an integrable logarithmic divergence for small argument, and decays with $e^{-2v}$ for large argument. Except for very small value of $v$, where the dominant contribution to $D_0$ is given by $1/\ln(v)$, $D_0$ can be well approximated by\footnote{In \cite{BA}, B. Allen {\it at all} adopted a similar procedure to provide an approximate expression to the vacuum expectation value of the field square in the cosmic string spacetime considering a general structure to its core; in \cite{Spinelly} this procedure has is also been adopted to calculate the $\langle\phi^*\phi\rangle$.} $-1/\ln(qr)$. In figure $1$, this argument is also numerically justified. There, it is exhibited the behavior of the exact integrand of $S(\zeta)$, and its approximated expression discarding the factor $\ln(v)$ in the denominator, for specific values of the parameters. 
\begin{figure}[tbph]
\begin{center}
\includegraphics[angle=90,angle=90,angle=90,width=9.5cm,height=7.cm]{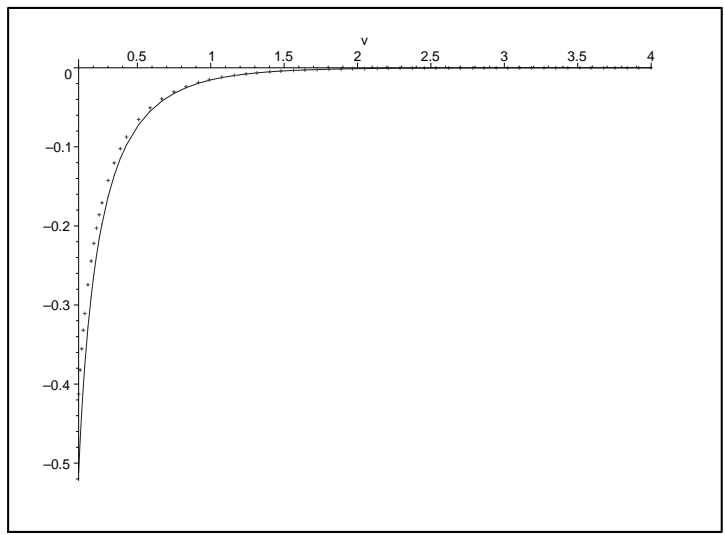}
\end{center}
\caption{The dashed curve corresponds the exact integrand of the second term of the right hand side of (\ref{j3}), and the solid line the approximated expression discarding $\ln(v)$ in the denominator of (\ref{jj3}). In the numerical analysis we have used $\alpha=0.99$, $\gamma=0.2$, $N=0$ and $r/R=10^3$.}
\label{fig1}
\end{figure}

Accepting the above arguments, we can represent the main contribution to (\ref{jj3}) by:
\begin{equation}
\bar{S}(\zeta)=-\frac{1}{\zeta \ln(qr)}\int_0^{\infty} dv   K^{2}_{0}(v) \ ,
\end{equation}
Consequently we obtain
\begin{equation}
\langle\hat{\phi}^{\ast}(x)\hat{\phi}(x)\rangle_{T}^{C}=-\frac{1}{8\alpha \beta r \ln \left[ \frac{2r}{R}e^{-\mathcal{C}+\alpha/\delta} \right]}  \ .
\end{equation}

Now for this model, the expression to the renormalized vacuum polarization effect in high-temperature limit reads
\begin{equation}
\langle\hat{\phi}^{\ast}(x)\hat{\phi}(x)\rangle_{T, Ren.}\approx \frac{1}{12 \beta^{2}}+\frac{M^{(\gamma)}}{\beta r}-\frac{1}{8\alpha \beta r \ln \left[ \frac{2r}{R}e^{-\mathcal{C}+\alpha/\delta} \right]} \ .
\end{equation}

On basis of previous discussion about the vacuum polarization effect at zero temperature, we can conclude that although being sub-dominant, the thermal correction due to the non vanishing radius of the magnetic flux tube, is so relevant as the standard term proportional to $1/\beta r$, for points at very large distance from the cosmic string. Consequently, it can be considered as a long-range effect.


\section{The Computation of $\langle\hat{T}_\mu^\nu(x)\rangle_{Ren.}$ at Non-zero Temperature}
\label{sec4}

The energy-momentum tensor, $T_{\mu\nu}(x)$, is a bilinear function of the fields, so we can evaluate its vacuum expectation value, 
$\langle T_{\mu\nu}(x)\rangle$, by the standard method using the Green function \cite{BD}. The thermal vacuum average, consequently, can also be obtained by using the thermal Green function.

The renormalized vacuum expectation value of the energy-momentum tensor at non-zero temperature for the system adopted here can be calculated by:
\begin{equation}
\label{EMT}
\langle\hat{T}_\mu^\nu\rangle_{T,Ren.}=\lim_{x'\to x}[D_{\mu(\alpha,\phi)}^{\nu'}G_{T}(x,x')-D_{\mu(1,0)}^{\nu'}G_{H}(x,x')] \ ,
\end{equation}
where $D_{\mu(\alpha,\phi)}^{\nu'}$ stands a second order non-local differential operator in presence of a magnetic field and cosmic string, defined by 
\begin{equation}
D_{\mu(\alpha,\phi)}^{\nu'}=(1-2\xi)D_\mu{\overline{D}}^{\nu'}-\xi (D_\mu {D}^\nu+\overline{D}_{\mu'}\overline{D}^{\nu'})+\left( 2\xi-\frac12 \right)\delta_\mu^\nu {D}_{\sigma}\overline{D}^{\sigma'} \ ,
\end{equation}
$\xi$ being the non-minimal curvature coupling. In the above expression we consider that ${D}_\sigma={\nabla}_\sigma- ieA_\sigma$, with ${\nabla}_\sigma$ being the covariant gravitational derivative and the bar denoting complex conjugate. Again we shall consider the magnetic field configuration given by (\ref{j9}) and (\ref{j10}) for the two first models, and by (\ref{jj10}) for the third model. The respective thermal Green function, $G_T(x',x)$, are given by (\ref{j0}) and (\ref{jj1}), respectively.

Due to the form of the Green function, we can express (\ref{EMT}) by
\begin{equation}
\langle\hat{T}_\mu^\nu(x)\rangle_{T,Ren.}=\langle\hat{T}_\mu^\nu(x)\rangle_{T,Reg.}+\langle\hat{T}_\mu^\nu(x)\rangle_{T=0}^{C}+\langle\hat{T}_\mu^\nu(x)\rangle_{T}^{C} \ .
\label{je}
\end{equation}

The first and second term of the right hand side of the above expression have already been computed in \cite{E} and \cite{Spinelly}, respectively. As to the first term, Guimar\~aes has shown that
\begin{eqnarray}
	\langle\hat{T}_\mu^\nu(x)\rangle_{T,Reg.}=\frac{\pi^2}{45\beta^4}diag(-3,1,1,1)+\langle{\tilde{T}}_\mu^\nu(x)\rangle_T \ ,
\end{eqnarray}
where the second term on the right hand side is given in terms of five integrals. However, the author was able to show that in the high-temperature limit ($\beta\to 0$), this term is proportional to $1/\beta r^3$, and in the zero-temperature limit ($\beta\to\infty$) $\langle\hat{T}_\mu^\nu(x) \rangle_{T,Ren.}$ is proportional to $1/r^4$. The second term in (\ref{je}), $\langle\hat{T}_\mu^\nu(x)\rangle_{T=0}^{C}$, is consequence of a non-vanishing radius to the magnetic flux. For the two first models, its most important contribution is proportional to $(1/r^4)(R/ r)^{2\gamma/\alpha}$, consequently is relevant only in the region near the magnetic tube; on the other hand, for the third model this term presents a long-range effect, similar to what happens in the vacuum expectation value of the field square. The new contribution, $\langle\hat{T}_\mu^\nu(x)\rangle_{T}^{C}$, is consequence of a non-vanishing radius attributed to the magnetic flux tube and the temperature. 

For the two first models this term is given by
\begin{equation}
\langle\hat{T}_\mu^\nu\rangle_{T}^C=\lim_{x'\to x}D_{\mu(\alpha,\phi)}^{\nu'}G_{T}^{C,(j)}(x,x') \ ,
\label{38}
\end{equation}
with
\begin{eqnarray}
G_{T}^{C,(j)}(x,x')&=&\frac{1}{4 \pi^{2} \alpha}\sum_{l \neq 0}\int_0^{\infty} d\omega \omega J_{0}\left(\omega\sqrt{(\Delta 
\tau+l\beta)^{2}+(\Delta z)^{2}}\right)\times
\nonumber\\
&&\sum_{n=-\infty}^{\infty}e^{i n\Delta\theta}D^j_n(\omega R)K_{|\nu|}(\omega r)K_{|\nu|}(\omega r') \  , \qquad{j=1,2} \ .
\label{39}
\end{eqnarray}

Following the same procedure adopted in the last section, we shall consider only the $n=N$ component in the summation. Substituting (\ref{39}) into (\ref{38}), and using the approximated expression to the coefficient $D^j_N$, after long calculation we arrive at the following  result:
\begin{eqnarray}
\langle\hat{T}_\mu^\nu\rangle_{T}^{C}&=&-\frac{1}{4 \pi^{2} \alpha r^{4}}\frac{2}{\Gamma\left( \gamma/\alpha \right)\Gamma\left( 1+\gamma/\alpha\right)}\left[ \frac{w_j^{N}-(\gamma/\alpha)z_j^{N}}{w_j^{N}+(\gamma/\alpha)z_j^{N}}\right]\left( \frac{R}{2r}\right)^{2\gamma/\alpha} \times \nonumber \\
&&\hbox{diag}\left(a_{0}^{(\gamma)},a_{1}^{(\gamma)},a_{2}^{(\gamma)},a_{3}^{(\gamma)}\right)
\end{eqnarray}
where
\begin{eqnarray}
a_{0}^{(\gamma)}=\left( 4\xi-1\right)\left[ I_{3}^{(\gamma)}+I_{1}^{(\gamma)}-2(\gamma/\alpha)I_{2}^{(\gamma)}+2\left( \gamma/\alpha \right)^{2}I_{4}^{(\gamma)} \right] \nonumber \ ,
\end{eqnarray}
\begin{eqnarray}
a_{1}^{(\gamma)}= I_{1}^{(\gamma)}-2\left(\gamma/\alpha+2\xi\right)I_{2}^{(\gamma)}+4\xi(\gamma/\alpha)I_{4}^{(\gamma)}-I_{3}^{(\gamma)}\nonumber \ ,
\end{eqnarray}
\begin{eqnarray}
a_{2}^{(\gamma)}&=&2(\gamma/\alpha)\left( \gamma/\alpha-2\xi\right)I_{4}^{(\gamma)}+4\xi I_{2}^{(\gamma)}+ \nonumber \\
&&\left( 4\xi-1\right)\left[I_{3}^{(\gamma)}+I_{1}^{(\gamma)}-2(\gamma/\alpha)I_{2}^{(\gamma)}+2\left( \gamma/\alpha \right)^{2}I_{4}^{(\gamma)} \right] \nonumber \ , 
\end{eqnarray}
\begin{eqnarray}
a_{3}^{(\gamma)}=2I_{5}^{(\gamma)}+\left( 4\xi-1\right)\left[I_{3}^{(\gamma)}+I_{1}^{(\gamma)}-2(\gamma/\alpha)I_{2}^{(\gamma)}+2\left( \gamma/\alpha \right)^{2}I_{4}^{(\gamma)} \right] \nonumber \ ,
\end{eqnarray}
with
\begin{eqnarray}
I_{1}^{(\gamma)}=\sum_{l=1}^{\infty}\int_{0}^{\infty} dv v^{3+2\gamma/\alpha}J_{0}\left( v \zeta l\right)K_{\gamma/\alpha+1}^{2}(v) \ ,
\label{I1}
\end{eqnarray}
\begin{eqnarray}
I_{2}^{(\gamma)}=\sum_{l=1}^{\infty}\int_{0}^{\infty} dv v^{2+2\gamma/\alpha}J_{0}\left( v \zeta l\right)K_{\gamma/\alpha+1}K_{\gamma/\alpha}(v) \ ,
\label{I2}
\end{eqnarray}
\begin{eqnarray}
I_{3}^{(\gamma)}=\sum_{l=1}^{\infty}\int_{0}^{\infty} dv v^{3+2\gamma/\alpha}J_{0}\left( v \zeta l\right)K_{\gamma/\alpha}^{2}(v) \ ,
\label{I3}
\end{eqnarray}
\begin{eqnarray}
I_{4}^{(\gamma)}=\sum_{l=1}^{\infty}\int_{0}^{\infty} dv v^{1+2\gamma/\alpha}J_{0}\left( v \zeta l\right)K_{\gamma/\alpha+1}^{2}(v) \ 
\label{I4}
\end{eqnarray}
and
\begin{eqnarray}
I_{5}^{(\gamma)}=\sum_{l=1}^{\infty}\int_{0}^{\infty} dv v^{2+2\gamma/\alpha}\frac{J_{1}\left( v \zeta l\right)}{\zeta l}K_{\gamma/\alpha}^{2}(v) \ .
\label{I5}
\end{eqnarray}
In order to provide the most relevant contribution to (\ref{I1})-(\ref{I4}), we adopt the same procedure adopted in last section: we divide the interval of integration in two parts, from $\left[0,2\pi/\zeta \right]$ and from $\left[2\pi/\zeta,\infty \right)$. Again in the high-temperature regime, $\zeta\ll 1$, the integral in the second interval can be neglected due the exponential decay of the modified Bessel function. Finally using the series properties for the Bessel function \cite{Grad}, and after some intermediate steps, we obtain:
\begin{eqnarray}
\langle\hat{T}_\mu^\nu\rangle_{T}^{C}=-\frac{B_{j}^{\gamma, \alpha, R}}{\beta r^{3}}\left( \frac{R}{r}\right)^{2 \gamma/\alpha}\hbox{diag}\left( b_{0}^{(\gamma)},b_{1}^{(\gamma)},b_{2}^{(\gamma)},b_{3}^{(\gamma)} \right) \  , \quad{j=1,2} \ ,
\end{eqnarray}
where
\begin{eqnarray}
B_{j}^{\gamma, \alpha, R}=\frac{\left( 1+2\gamma/\alpha \right)}{2\pi^{3/2}\alpha}\frac{\Gamma^{2}\left( 1/2+\gamma/\alpha\right) \Gamma\left(1/2+2\gamma/\alpha \right)}{\Gamma\left(3+2\gamma/\alpha \right)\Gamma(1+\gamma/\alpha)\Gamma(\gamma/\alpha)}\left[\frac{w^N_j
-(\gamma/\alpha)z^{N}_{j}}{w^{N}_{j}+(\gamma/\alpha)z^{N}_{j}}\right]  \ , \nonumber
\end{eqnarray}
being
\begin{eqnarray}
b_{0}^{(\gamma)}=\left( \xi -\frac{1}{4}\right)\left[1+5(\gamma/\alpha)+8\left(\gamma/\alpha \right)^{2}+4\left(\gamma/\alpha \right)^{3} \right] \nonumber \ ,
\end{eqnarray} 
\begin{eqnarray}
b_{1}^{(\gamma)}=\frac{1}{8}\left\{1+4(\gamma/\alpha)-8\xi\left[1+3(\gamma/\alpha)+2\left( \gamma/\alpha \right)^{2}\right] \right\} \nonumber \ ,
\end{eqnarray} 
\begin{eqnarray}
b_{2}^{(\gamma)}=-\frac{1}{4}\left\{4+5(\gamma/\alpha)-4\left( \gamma/\alpha\right)^{2}-8\xi\left[1+4(\gamma/\alpha)+5\left( \gamma/\alpha\right)^{2} +2\left( \gamma/\alpha\right)^{3}\right] \right\} \nonumber 
\end{eqnarray}
and
\begin{eqnarray}
b_{3}^{(\gamma)}&=&-\frac{1}{8}\left\{ 1+4(\gamma/\alpha)+8\left( \gamma/\alpha\right)^{2}+8\left(\gamma/\alpha\right)^{3}\right. \nonumber \\
&& \left. -8\xi\left[1+5(\gamma/\alpha)+8\left( \gamma/\alpha\right)^{2}+4\left( \gamma/\alpha\right)^{3}\right]  \right\} \nonumber \ .
\end{eqnarray}

Analysing the result we can observe that $\langle\hat{T}_\mu^\nu\rangle_{T}^{C}$ becomes relevant in the region near the magnetic tube. In this region, it is of the same order of magnitude as the sub-dominant contribution obtained in \cite{E} in the high-temperature limit. Also it is possible to check that for the conformal factor $\xi=1/6$, the trace of the correction vanishes, i.e., $\langle\hat{T}_\mu^\mu\rangle_{T}^{C}=0$

For the third model the new contribution is given by:
\begin{equation}
\langle\hat{T}_\mu^\nu\rangle_{T}^C=\lim_{x'\to x}D_{\mu(\alpha,\phi)}^{\nu'}G_{T}^{C,(3)}(x,x') \ ,
\label{je1}
\end{equation}
where
\begin{eqnarray}
G_{T}^{C,(3)}(x,x')&=&\frac{1}{4 \pi^{2} \alpha}\sum_{l \neq 0}\int_0^{\infty} d\omega \omega J_{0}\left(\omega\sqrt{(\Delta \tau+l\beta)^{2}+(\Delta z)^{2}}\right)\times
\nonumber\\
&&\sum_{n=-\infty}^{\infty}e^{i n\Delta\theta}D_n(\omega R)K_{|n|/\alpha}(\omega r)K_{|n|/\alpha}(\omega r') \ .
\label{je2}
\end{eqnarray}

Using the same procedure adopted to calculate the thermal average of the field square, we shall consider only the component $n=0$ in the thermal Green function above. Substituting (\ref{je2}) into (\ref{je1}), using the approximated expression to the coefficient $D_0$, and taking into account the same considerations to overcome the integral problem found in (\ref{jj3}), we obtain:
\begin{eqnarray}
\langle\hat{T}_\mu^\nu\rangle_{T,Ren.}=-\frac{1}{4\pi^{2}\alpha r^{4} \ln{\left(q/r\right)}}\hbox{diag} \left(a_{0}^{(0)},a_{1}^{(0)},a_{2}^{(0)},a_{3}^{(0)} \right) \ ,
\end{eqnarray}
where the coefficients $a_\mu^{(0)}$ are given by the expressions found in the precedent analysis by taking $\gamma=0$. Moreover in the high-temperature regime ($\zeta \ll 1$), we obtain:
\begin{eqnarray}
\langle\hat{T}_\mu^\nu\rangle_{T,Ren.}=-\frac{1}{8 \alpha \beta r^{3} \ln{\left[ \frac{2r}{R} e^{-\mathcal{C}+\alpha/\delta}\right]}}\hbox{diag}(b_{0}^{(0)},b_{1}^{(0)},b_{2}^{(0)},b_{3}^{(0)}) \ .
\end{eqnarray} 
Here we also observe that a long-range effect appears, consequently it is so relevant as the sub-dominant contribution proportional to $1/\beta r^3$ for points at large distance from the tube.

\section{Concluding Remarks}
\label{concluding}
In this paper we have analysed the thermal effects on the renormalized vacuum expectation value of the square of a charged massless scalar field, $\langle\hat{\phi}^{\ast}\hat{\phi}\rangle$, and in the energy-momentum tensor, $\langle\hat{T}_\mu^\nu\rangle$, in the cosmic string spacetime considering the presence of a magnetic flux of finite radius. Three specific configurations of magnetic filed have been considered. For them we could express these vacuum polarization effects as the sum of three different terms as follows:
\begin{itemize}
\item The first one represents the thermal contribution coming from the interaction between the charged field with a magnetic flux considered as a line running along the cosmic string. This contribution has been exactly calculated in \cite{E}. 
\item The second term is the zero temperature contribution on the vacuum polarization due to the non vanishing radius of the magnetic flux tube. This contribution has been analysed in detail in \cite{Spinelly}. 
\item The third contribution is the new one. It comes from the combination of the non vanishing radius of the magnetic flux tube and temperature. It goes to zero for $R\to 0$ and for $T\to 0$. 
\end{itemize}

Unfortunately this new contribution cannot be expressed in terms of any analytical function. In order to provide the most important quantitative information about the thermal behavior of this contribution, we adopted an approximated procedure. We consider this term in the high-temperature regime. Doing this it was possible, by using the series property for the Bessel function, to obtain an analytical expression to this correction. 

Apart from the homogeneous thermal contribution to the vacuum polarizations, there appears in calculations of the thermal average of the field square and energy-momentum tensor, corrections due to the geometry of the spacetime, and due to the non-vanishing magnetic flux. The latter presents two parts: one is given as the magnetic flux be a line running along the string, the standard contribution, and the other consequence of a finite transversal radius. These corrections are sub-dominant and depend on the distance to the string. For the two first models for the magnetic fields, the corrections on the thermal average due to the finite radius, are very similar and only relevant in the regions near the tube; as to the third model, it becomes so important as the standard one for large distance.  

Although the structure of the magnetic field produced by a $U(1)-$gauge cosmic string cannot be presented by any analytical function, its influence on the thermal vacuum polarization effects of charged matter fields takes place for sure. So in this case, the geometric and magnetic interactions provide contributions. The relevant physical question is how important they are. Trying to clarify this important question, we adopted specific fields configurations which allow us to develop an analytical procedure: we assumed that the magnetic field extent $R$ is much bigger that the cosmic string's radius considered here equal to zero.\footnote{In fact for the $U(1)-$gauge cosmic string the magnetic field extent is approximately $\delta\approx\frac1m\frac{\sqrt{\lambda}}e$ \cite{N-O}, while the cosmic string radius $\xi\approx\frac1m$. So we are considering the case where $\frac{\sqrt{\lambda}}e>>1$.} So our main conclusion is that: although the structure of the magnetic field cannot be very well understood, its influence on the thermal vacuum polarization effect can be so relevant as the influence of the gravitational filed itself.  

A real cosmic string would be immersed in a bath of primordial heat radiation. The behavior of particle detectors near straight strings immersed in thermal radiation, in free space or passing through black holes, has been analyzed in \cite{Davies}. The influence of a magnetic field surrounding the string on this detector, can be evaluated by using the corresponding thermal Green functions calculated in this paper. In addition, some results obtained here may shed light upon the vacuum polarization effects induced by a realistic vortex configuration in early cosmology, where the temperature of the Universe was really high. By these results, we can see that thermal contributions to the vacuum expectation value of the field square and the energy-momentum tensor, modify the zero-temperature quantities in that epoch. Because of this, they should be taken into account in, for example, the theory of structure formation.
\\       \\

{\bf{Acknowledgment}}
\\       \\
One of us (ERBM) wants to thanks Conselho Nacional de Desenvolvimento Cient\'\i fico e Tecnol\'ogico (CNPq.) for partial financial support, FAPESQ-PB/CNPq. (PRONEX) and FAPES-ES/CNPq. (PRONEX).


\begin{thebibliography}{99}
\bibitem{Kibble}Kibble T W, J. Phys. {\bf A 9}, 1387 (1976).
\bibitem{Vilenkin}A. Vilenkin, Phys. Rep., {\bf 121}, 263 (1985).
\bibitem{N-O} N. B. Nielsen and P. Olesen, Nucl. Phys. {\bf B61}, 45 (1973).
\bibitem{Garfinkle} D. Garfinkle, Phys. Rev. D {\bf 32}, 1323 (1985).
\bibitem{Linet} B. Linet, Phys. Lett. B {\bf 124}, 240 (1987).
\bibitem{Yves} M. Christensen, A. L. Larsen and Y. Verbin, Phys. Red. D {\bf 60}, 125012 (1999); Y. Brihaye and M. Lubo, {\it ibid} {\bf 62}, 085005 (2000).
\bibitem{1Spinelly} J. Spinelly and E. R. Bezerra de Mello, Int. J. Mod. Phys. A {\bf 17}, 4375 (2002).
\bibitem{Spinelly} J. Spinelly and E. R. Bezerra de Mello, Class. Quantum Grav. {\bf 20}, 873 (2003).
\bibitem{2Spinelly} J. Spinelly and E. R. Bezerra de Mello, Int. J. Mod. Phys. D {\bf 13}, 607 (2004). 
\bibitem{3Spinelly} J. Spinelly and E. R. Bezerra de Mello, Nucl Phys. B (Proc. Suppl.) {\bf 127}, 77 (2004).
\bibitem{S} E. M. Serebryanyi, Theor. Math. Phys. {\bf 64}, 846 (1985).
\bibitem{E-L} M. E. X. Gumar\~aes and B. Linet, Commun. Math. Phy. {\bf 165}, 297 (1994).
\bibitem{E} M.E.X. Guimar\~aes, Class. Quantum Grav. {\bf 12}, 1705 (1995).
\bibitem{Mello} F. C. Cabral and E. R. Bezerra de Mello, Class. Quantum Grav. {\bf 18}, 1637 (2001).
\bibitem{Mello1} F. C. Cabral and E. R. Bezerra de Mello, Class. Quantum Grav. {\bf 18}, 5455 (2001).
\bibitem{Mello2} E. R. Bezerra de Mello and F. C. Cabral, Int. J. Mod. Phys. A {\bf 17} 879 (2002).
\bibitem{Grad} I. S. Gradshteyn e I. M Ryzhik, {\it{Table of Integral, Series and Products}} (Academic Press, New York, 1980).
\bibitem{smith} A. G. Smith, in {\it{Symposium on the Formation and Evolution of Cosmic String}}, edited by G. W. Gibbons, S. W. Hawking and T. Vachaspati (Cambridge University Press, Cambridge, England, 1989).
\bibitem{linet2} B. Linet, Phys. Rev. D, {\bf{35}}, 536 (1987).
\bibitem{Braden}H. W. Braden, Phys. Rev. D {\bf 25}, 1028 (1982).
\bibitem{Page}D. N. Page, Phys. Rev. D {\bf 25}, 1499 (1982).
\bibitem{linet1} B. Linet, Class. Quantum Grav. {\bf 9}, 2429 (1992).
\bibitem{BA} B. Allen, B. S. Kay and A. C. Ottewill, Phys. Rev. D {\bf 53}, 6829 (1996).
\bibitem{Spi} J. Spinelly and E. R. Bezerra de Mello, PoS(IC2006), 056 (2006).
\bibitem{BD} N. D. Birrell and P. C. W Davis, {\it Quantum Fields in Curved Space} (Cambridge University Press, Cambridge, England, 1982).
\bibitem{Davies} P. C. Davies and V. Sahni, Class. Quantum Grav. {\bf 5}, 1 (1988).


\end{thebibliography}
\end{document}